\documentclass[aps,prd,showpacs,twocolumn,floatfix,nofootinbib]{revtex4-1}
\usepackage[dvipdfmx]{graphicx}
\usepackage{amsmath,amssymb,siunitx}
\usepackage{color,ulem}
\usepackage{graphicx}
\usepackage{bm,latexsym,amsmath,amssymb,amsfonts,mathrsfs}
\usepackage{dcolumn}   
\usepackage{color}

\newcommand{\simgt}{\lower.5ex\hbox{$\; \buildrel > \over \sim \;$}}
\newcommand{\simlt}{\lower.5ex\hbox{$\; \buildrel < \over \sim \;$}}

\begin{document}
\title{Gravitational-wave constraints on the GWTC-2 events by measuring the tidal deformability and the spin-induced quadrupole moment}
\author{Tatsuya Narikawa$^{1}$}
\email{narikawa@icrr.u-tokyo.ac.jp}
\author{Nami Uchikata$^{1}$}
\email{uchikata@icrr.u-tokyo.ac.jp}
\author{Takahiro Tanaka$^{2,3}$}
\email{t.tanaka@tap.scphys.kyoto-u.ac.jp}
\affiliation{
$^1$Institute for Cosmic Ray Research, The University of Tokyo, Chiba 277-8582, Japan\\
$^2$Department of Physics, Kyoto University, Kyoto 606-8502, Japan\\
$^3$Center for Gravitational Physics, Yukawa Institute for Theoretical Physics, Kyoto University, Kyoto 606-8502, Japan\\
}
\date{\today}

\begin{abstract}
Gravitational waves from compact binary coalescences provide a unique laboratory to test properties of compact objects. 
As alternatives to the ordinary black holes in general relativity, various exotic compact objects have been proposed. 
Some of them have largely different values of the tidal deformability and spin-induced quadrupole moment 
from those of black holes, and their binaries could be distinguished from the binary black hole by using gravitational waves 
emitted during their inspiral regime, excluding the highly model-dependent merger and ringdown regimes. 
We reanalyze gravitational waves from low-mass merger events in the GWTC-2, 
detected by the Advanced LIGO and Advanced Virgo.
Focusing on the influence of tidal deformability and spin-induced quadrupole moment in the inspiral waveform, 
we provide model-independent constraints on deviations from the standard binary black hole case. 
We find that all events that we have analyzed are consistent with the waveform of the binary black hole in general relativity. Bayesian model selection shows that the hypothesis that the binary is composed of exotic compact objects 
is disfavored by all events.
\end{abstract}


\maketitle

\section{Introduction}
\label{sec:Introduction}
Gravitational waves (GWs) that are thought to originate from binary black hole (BBH) coalescences 
have been detected with the Advanced LIGO and Advanced Virgo interferometers~\cite{LIGOScientific:2018mvr, Abbott:2020niy}.
General relativity (GR) has been tested using GWs from compact binary coalescences in several ways~\cite{Abbott:2020jks}.

Exotic compact objects (ECOs), such as boson stars, gravastars, wormholes, 
and some models of quantum corrections at around the horizon, 
have been proposed as alternatives to the black holes (BHs) of classical GR (see Ref.~\cite{Cardoso:2019rvt} for review).
Some proposed ECOs have largely different values of the tidal deformability and spin-induced quadrupole moment (SIQM) 
from those of BHs in GR.
The inspiral regime for GW signals emitted from binary ECOs can be characterized through  
the tidal deformability~\cite{PoissonWillGraivty, Flanagan:2007ix, Hinderer:2007mb, Damour:2012yf, Agathos:2015uaa}
and SIQM~\cite{PoissonWillGraivty, Poisson:1997ha}.
Therefore, their binaries could be distinguished from BBHs by using GWs.

Tidal deformability and SIQM can also characterize the equation-of-state (EOS) for a neutron star (NS).
For the binary-neutron-star mergers, 
most analyses of GW170817 and GW190425 focus on measuring the tidal deformability 
using the EOS-insensitive relations between tidal deformability and SIQM called Love-Q relations predicted 
by theoretical studies~\cite{Yagi:2013bca, Yagi:2013awa, Abbott:2018wiz, Abbott:2018exr, Abbott:2020uma, Narikawa:2018yzt, Narikawa:2019xng}.

Regarding the tidal absorption effects~\cite{Hartle:1973zz, Hughes:2001jr, Maselli:2017cmm, Isoyama:2017tbp},
some studies have proposed that some ECOs such as boson stars do not absorb GWs efficiently~\cite{Maselli:2017cmm, Datta:2019epe, Datta:2019euh, Datta:2020gem} (see Ref.~\cite{Cardoso:2019rvt} for review).
In this paper, we ignore the tidal absorption effects because of uncertainty of theoretical understanding, 
e.g., uncertainty of frequency dependence on horizon energy flux~\cite{Sago:2021iku}.

At the very late inspiral regime, the tidal resonance for binary ECO systems is also modeled in Ref.~\cite{Fransen:2020prl},
and resonant excitations in ECOs 
have been constrained via GWTC-1 events~\cite{Asali:2020wup}.
At postmerger regimes for binary ECO, emission of GW echoes following normal inspiral-merger-ringdown signal 
have been proposed and have been searched~\cite{Abedi:2016hgu, Westerweck:2017hus, Uchikata:2019frs}.
A toy model for postcontact regimes of binary ECOs has been recently introduced~\cite{Toubiana:2020lzd}.
In this paper, we do not consider the postinspiral regimes, such as the resonance and GW echoes.

Previous studies regarding real data analysis of GWs to test ECO hypothesis 
have considered only the effect of either the tidal deformability or the SIQM on the signal,
and the BH's value for the other parameter has been used.
The studies focusing on the SIQM effect alone have been done in Refs.~\cite{Kastha:2018bcr, Krishnendu:2017shb, Krishnendu:2018nqa, Krishnendu:2019ebd, Krishnendu:2019tjp}.
In particular, Krishnendu {\it et al.} have constrained the properties of GW151226 and GW170608 
from the measurements of SIQM~\cite{Krishnendu:2019tjp}, and 
the paper on tests of GR by the LIGO Scientific Collaboration and the Virgo Collaboration (LVC) has constrained 
the ECO hypothesis with the GWTC-2 events~\cite{Abbott:2020jks}.
The studies focusing on the tidal effect alone have been done in Refs.~\cite{Wade:2013hoa, Cardoso:2017cfl, Sennett:2017etc, Johnson-McDaniel:2018uvs}.
In particular, Johnson-Mcdaniel {\it et al.} have tested how well BH mimickers with polytropic EOS are constrained
by measuring tidal deformabilities via injected O1 BBH-like evetns~\cite{Johnson-McDaniel:2018uvs}.
They show future prospects of detection of deviations from GR assuming boson stars.
We note, however, that the above works do not consider both effects simultaneously but focus
on only one of them.\footnote{The analysis in Ref.~\cite{Pacilio:2020jza} has considered both effects at 
one time for Fisher information matrix analysis.}

The aim of this work is to give model-independent tests of strong-field gravity regimes 
from the measurements of tidal deformability and SIQM via GWs from compact binary inspirals. 
One motivation to think about BH mimickers is to modify the BHs in GR to be compatible 
with the stringy paradigm as to BH information loss. 
If we assume something like a firewall, only the small region near the horizon might be modified, 
or in other words practically only the absorbing boundary condition across the horizon might be modified. 
This change of boundary condition may also result in the modification of tidal deformability, 
unless the modification is restricted to a really tiny region in the vicinity of the horizon, 
e.g., within the Planck distance from the horizon. 
(Thus, we think nontrivial tidal deformability will not necessarily imply smaller compactness.)

In this paper, we reanalyze the data around six low-mass events identified as BBH; 
GW151226,  GW170608, GW190707\_093326 (hereafter GW190707), 
GW190720\_00836 (hereafter GW190720), GW190728\_064510 (hereafter GW190728), 
and GW190924\_021846 (hereafter GW190924),
using an inspiral-only waveform model with both tidal and SIQM terms,
and present constraints on the binary tidal deformability and SIQM at the same time.
We focus on the inspiral regime because postinspiral regimes of binary ECOs are not modeled well.
Since the inspiral regime can be accurately described 
by the post-Newtonian (PN) formula~\cite{Blanchet:2013haa, PoissonWillGraivty, Isoyama:2020lls}, 
we use the PN inspiral-only waveform model.

The remainder of this paper is organized as follows.
In Sec.~\ref{sec:PE}, we explain the methods of Bayesian parameter estimation for GWs 
including waveform models used to analyze.
In Sec.~\ref{sec:results}, we present results of our analysis of GWTC-2 events 
by using \texttt{TF2g\_Tidal\_SIQM}~waveform model.
Section~\ref{sec:summary} is devoted to a summary and conclusion.
In Appendix~\ref{sec:results_TF2}, we show the results for seven events added GW190814 
by using the \texttt{TF2\_Tidal\_SIQM}~waveform model.
Here, \texttt{TF2}~is the abbreviation of \texttt{TaylorF2}, 
which is the PN waveform model for point-particle and spin effects~\cite{Dhurandhar:1992mw, Buonanno:2009zt, Blanchet:2013haa},
and \texttt{TF2g} is an extended waveform model of \texttt{TF2}
obtained by Taylor-expanding the effective-one-body formula~\cite{Messina:2019uby}.

We employ the units $c=G=1$, where $c$ and $G$ are the speed of light 
and the gravitational constant, respectively.

\section{Parameter estimation methods}
\label{sec:PE}
\subsection{ECO features}
\label{sec:features}
There are several features of ECOs that differ from BH (see Refs.~\cite{Cardoso:2019rvt} for review).
In this paper, we focus on the tidal deformability and SIQM.

\subsubsection{Tidal deformability}
\label{sec:tidal}
In the compact binary inspiral, at the leading order,
the tidally induced quadrupole moment tensor $Q_{ij,\mathrm{Tidal}}$ is proportional to 
the companion's tidal field ${\cal E}_{ij}$ as $Q_{ij,\mathrm{Tidal}} = -\lambda {\cal E}_{ij}$.
The information about the EOS (or structure) can be quantified by the tidal deformability parameter 
$\lambda$~\cite{Flanagan:2007ix}.
The leading-order tidal contribution to the GW phase evolution (relative 5PN order) arises
through the symmetric contribution of  tidal deformation, the binary tidal deformability~\cite{Flanagan:2007ix, Hinderer:2007mb, Vines:2011ud}
\begin{eqnarray}
 \tilde{\Lambda} = \frac{16}{13} \frac{(m_1+12m_2)m_1^4\Lambda_1+(m_2+12m_1)m_2^4\Lambda_2}{(m_1+m_2)^5},
\end{eqnarray}
which is a mass-weighted linear combination of the both component tidal parameters,
where $m_{1,2}$ is the component mass and 
$\Lambda_{1,2}$ is the dimensionless tidal deformability parameter of each object 
defined as $\Lambda_{1,2}=\lambda_{1,2}/m_{1,2}^5$.
For the waveform models used in this paper, the tidal effects to the gravitational-wave phase 
are dominated by the symmetric contributions, $\tilde{\Lambda}$ terms, 
and the antisymmetric contributions, 
$\delta \tilde{\Lambda}$ terms, 
are always subdominant~\cite{Favata:2013rwa, Wade:2014vqa}.
The tidal deformability can characterize the compact objects.
$\Lambda$ for BHs in classical GR vanishes as shown for a Schwarzschild BH~\cite{Binnington:2009bb, Damour:2009wj}
and for Kerr BH~\cite{Poisson:2014gka, Pani:2015hfa, Landry:2015zfa}.
For NSs, $\Lambda$ is a few hundred, depending on the EOS~\cite{Hinderer:2009ca, Postnikov:2010yn, Vines:2011ud, Damour:2012yf, Lattimer:2015nhk}.
The upper bound on the binary tidal deformability $\tilde{\Lambda}$ by GW170817 is about 900~\cite{TheLIGOScientific:2017qsa, Abbott:2018wiz} (see also Refs.~\cite{Narikawa:2018yzt, Narikawa:2019xng} for reanalysis).
ECOs differ from BHs in classical GR in having nonzero tidal deformabilities~\cite{Cardoso:2017cfl, Sennett:2017etc, Cardoso:2018ptl}.
It is intriguing that $\Lambda$ is negative for gravastars~\cite{Pani:2015tga, Uchikata:2016qku}.

\subsubsection{SIQM}
\label{sec:SIQM}
For a compact object with mass $m$ and the dimensionless spin parallel to the orbital angular momentum, $\chi = S/m^2$, 
where $S$ is the magnitude of the spin angular momentum of the aligned component, 
the spin-induced quadrupole moment scalar is given by~\cite{Poisson:1997ha}
\begin{eqnarray}
 Q_\mathrm{Spin} = - (1 + \delta\kappa) \chi^2 m^3,
 \label{eq:SIQM}
\end{eqnarray}
where $\delta\kappa$ denotes deviations from the Kerr BHs in GR.
The symmetric combination of the deviation parameters of the respective objects, $\delta\kappa_{1,2}$, is defined as 
$\delta\kappa_s = (\delta\kappa_1 + \delta\kappa_2)/2$.
The SIQM can also characterize the compact objects.
For Kerr BH, we have $\delta\kappa=0$~\cite{Poisson:1997ha}.
For spinning NS, we have $\delta\kappa \sim 2-20$~\cite{Pappas:2012qg, Pappas:2012ns, Harry:2018hke}.
ECOs differ from BHs, having $\delta\kappa \neq 0$:
$\delta\kappa \sim 10-150$ for spinning boson stars~\cite{Ryan:1996nk, Ryan:1997hg, Herdeiro:2014goa, Baumann:2018vus}.
Interestingly, $\delta\kappa$ for a gravastar can take negative values~\cite{Uchikata:2016qku}.

\subsection{Waveform models for inspiraling exotic compact objects}
\label{sec:WF}

We use the PN waveform model, 
which can accurately describe the early inspiral regime for compact binary coalescences~\cite{Blanchet:2013haa, PoissonWillGraivty}.
The frequency-domain gravitational waveform for binary ECOs can be written as
\begin{eqnarray}
 \tilde{h}_\mathrm{ECO}(f) = A(f) e^{i[\Psi_\mathrm{BBH}(f) + \Psi_\mathrm{SIQM}(f) + \Psi_\mathrm{Tidal}(f)]},
\end{eqnarray}
where $A(f)$ is the amplitude of the GW signal and the phase which consists of the BBH phase $\Psi_\mathrm{BBH}(f)$ 
and the additional SIQM effect $\Psi_\mathrm{SIQM}(f)$ and tidal effect $\Psi_\mathrm{Tidal}(f)$.
We consider the amplitude formula up to the 3PN order as summarized in Ref.~\cite{Khan:2015jqa}, 
where the point particle and the spin effects are included for both BBH and ECO hypotheses
and the leading-order term is approximately written as approximately
$\sim d_\mathrm{L}^{-1} ({\cal M}^\mathrm{det})^{5/6} f^{-7/6}$
where $d_\mathrm{L}$ is the luminosity distance to the source
and 
${\cal M}^\mathrm{det} = (1+z) (m_1 m_2)^{3/5}/(m_1+m_2)^{1/5}$
is the detector-frame (redshifted) chirp mass, 
which gives the leading-order evolution of the binary amplitude and phase, 
and $z$ is the source redshift.

We use an extended model of \texttt{TaylorF2}~\cite{Dhurandhar:1992mw, Buonanno:2009zt, Blanchet:2013haa}, \texttt{TF2g},
as $\Psi_\mathrm{BBH}(f)$,
which consists of point-particle and spin parts.
For \texttt{TaylorF2}~the 3.5PN-order formulas is employed for the point-particle part of the phase 
as summarized in Refs.~\cite{Buonanno:2009zt, Khan:2015jqa}.
For \texttt{TF2g}, the phase of the point-particle part is extended to the quasi 5.5PN order, 
which is derived by the Taylor expansion of the effective-one-body 
formula taking into account the notion in the test particle limit~\cite{Messina:2019uby}.
We set the uncalculated terms at 4PN order and beyond to zero.
The added higher PN-order terms enable us to reduce the tidal deformability biasing.

The waveform models for both BBH and binary ECO used in our parameter estimation analyses
assume that the spins of component objects are aligned with the orbital angular momentum
and incorporate the 3.5PN-order formula in couplings between the orbital angular momentum 
and the component spins~\cite{Bohe:2013cla},
3PN-order formula in point-mass spin-spin, 
and self-spin interactions~\cite{Arun:2008kb, Mikoczi:2005dn} as summarized in Ref.~\cite{Khan:2015jqa}.
The different PN-order terms for spin effects could help to break degeneracy between parameters.

We also use the PN formula for the tidal effects.
We employ the 2.5PN-order (relative 5+2.5PN-order) tidal part of the phase.
Recently, Henry {\it et al.}~\cite{Henry:2020ski} have derived the complete form up to 5+2.5PN order
for the mass quadrupole, current quadrupole, and mass octupole contributing to the GW tidal phase.
We rewrite it only for the mass quadrupole interactions
as a function of the dimensionless tidal deformability of each object defined as $\Lambda_{1,2}=\lambda_{1,2}/m_{1,2}^5$ by\footnote{We will extend the our analysis by adding the current quadrupole and mass octupole interactions.}
\begin{widetext}
\begin{eqnarray}
 &&\Psi_\mathrm{Tidal}(f) = \frac{3}{128\eta} x^{5/2} \sum_{A=1}^{2} \Lambda_A X_A^4 
 \left[ -24(12-11 X_A) - \frac{5}{28} (3179-919 X_A - 2286 X_A^2 + 260 X_A^3 ) x \right. \nonumber \\
 && +24 \pi (12 - 11 X_A) x^{3/2} \nonumber \\
 && -5 \left( \frac{193986935}{571536} - \frac{13060861}{381024} X_A - \frac{59203}{378} X_A^2
 - \frac{209495}{1512} X_A^3 + \frac{965}{54} X_A^4 - 4 X_A^5 \right) x^2 \nonumber \\
 &&  \left. + \frac{\pi}{28} (27719 - 22415 X_A + 7598 X_A^2 - 10520 X_A^3) x^{5/2}  \right], 
\end{eqnarray}
\end{widetext}
where $x=[\pi M_\mathrm{tot} (1+z) f]^{2/3}$ is the dimensionless PN expansion parameter, 
$M_\mathrm{tot} = m_1+m_2$ is the total mass, $\eta=m_1 m_2 / (m_1+m_2)^2$ is the symmetric mass ratio, 
and $X_A=m_A/M_\mathrm{tot}$, A=1,2.\footnote{Following the modification of the tidal contributions to the phasing in Ref. \cite{Henry:2020ski_v4}, Eq. (4) has been corrected. Two coefficients of 7PN order have been corrected.}
Here, we do not ignore the antisymmetric contribution $\delta \tilde{\Lambda}$ terms,
while they are always subdominant on the tidal effects to the GW phase, compared with
the symmetric contribution $\tilde{\Lambda}$ terms~\cite{Favata:2013rwa,Wade:2014vqa}.
We have implemented the complete GW tidal phase up to 5+2.5 PN-order and the tidal amplitude up to 5+1 PN-order 
for the mass quadrupole interactions and used it in our analyses.

The leading-order effect due to the SIQM appears as 
a part of spin-spin interactions in the PN phase at relative 2PN order~\cite{Poisson:1997ha}, and 
the leading-order additional term for binary ECO is described as
\begin{eqnarray}
\Psi_\mathrm{SIQM}(f) = \frac{75}{64} \frac{\delta\kappa_1 m_1^2 \chi_1^2 + \delta\kappa_2 m_2^2 \chi_2^2}{m_1 m_2} x^{-1/2}.
\end{eqnarray}
In our analysis, we also incorporate relative 3PN corrections to the GW phase due to the SIQM effects as described in 
Refs.~\cite{Arun:2008kb, Mishra:2016whh, Krishnendu:2017shb}.
The expression for the SIQM parts in terms of $\delta\kappa_s$ and $\delta\kappa_a$ are described in Ref.~\cite{Krishnendu:2017shb},
in which $\delta\kappa_a$ are the antisymmetric combination of the component SIQMs
defined as $\delta\kappa_a=(\delta\kappa_1-\delta\kappa_2)/2$.

In summary, our template models are the \texttt{TF2g}, \texttt{TF2g\_Tidal}, \texttt{TF2g\_SIQM}, and \texttt{TF2g\_Tidal\_SIQM}~waveform models,
which are, respectively, the reference BBH model, the ones with only the tidal terms, the ones with only the SIQM terms, 
and the ones with both the tidal and the SIQM terms.

\subsection{Bayesian inference}
\label{sec:Bayesian}
We employ Bayesian inference for GW parameter estimation and model selection 
(see Refs.~\cite{Thrane:2018qnx, Veitch:2014wba} for review)
between binary ECO and BBH.
%
%
Given data $d$, which contains the signal and the noise, 
according to Bayes's theorem, 
the posterior distribution of the signal parameters $\theta$ that the waveform $\tilde{h}(\theta)$ depends on
is given by 
\begin{eqnarray}
 p(\theta | d) = \frac{{\cal L}(d | \theta)\pi(\theta)}{\cal Z},
\end{eqnarray}
where ${\cal L}(d | \theta)$ is the likelihood function of the data for given parameters $\theta$, 
$\pi(\theta)$ is the prior distribution for $\theta$,
and ${\cal Z}$ is the evidence.
By assuming stationarity and Gaussianity for the detector noise,
the likelihood function is evaluated as,
\begin{eqnarray}
 {\cal L}(d | \theta) \propto \exp \left[ - \frac{\langle d - h(\theta) | d - h(\theta) \rangle}{2}\right],
\end{eqnarray}
where the noise-weighted inner product $\langle \cdot | \cdot \rangle$ is defined by
\begin{eqnarray}
 \langle a | b \rangle := 4 \mathrm{Re} \int_{f_\mathrm{low}}^{f_\mathrm{high}} df \frac{\tilde{a}^*(f) \tilde{b}(f)}{S_n(f)},
\end{eqnarray}
using the noise power spectrum density $S_n(f)$. 
We use $S_n(f)$ obtained with the \texttt{BayesLine}~algorithm~\cite{Cornish:2014kda, Littenberg:2015kpb, Chatziioannou:2019zvs}.
The lower limit of the integration $f_\mathrm{low}$ is the seismic cutoff frequency
and the higher limit $f_\mathrm{high}$ is the cutoff frequency of waveforms.
To restrict the analysis to the inspiral regimes of the signals, we set the upper frequency cutoff $f_\mathrm{high}$ 
to be referred to the one used in Refs.~\cite{Abbott:2020jks, Uchikata:2021jmy}.

The evidence is obtained as the likelihood marginalized over the prior volume,
\begin{eqnarray}
 {\cal Z} = \int d\theta {\cal L}(d | \theta) \pi(\theta).
\end{eqnarray}
To perform model selection between the binary ECO and BBH hypotheses,
we compute the ratio between two different evidences, called the Bayes factor,
\begin{eqnarray}
 \mathrm{BF}_\mathrm{BBH}^\mathrm{ECO} = \frac{{\cal Z}_\mathrm{ECO}}{{\cal Z}_\mathrm{BBH}}.
\end{eqnarray}
The combined Bayes factor is defined as
\begin{eqnarray}
 \log_{10} \mathrm{BF}_\mathrm{BBH,total}^\mathrm{ECO} = \sum_i \log_{10} \mathrm{BF}_\mathrm{BBH,i}^\mathrm{ECO},
\label{eq:combinedBF}
\end{eqnarray}
where $\mathrm{BF}_\mathrm{BBH,i}^\mathrm{ECO}$ is the Bayes factor for individual events.
The one-dimensional posterior for a specific parameter is obtained 
by marginalizing the multidimensional posterior over the other parameters.

We compute posterior probability distribution functions (PDFs)
by using Bayesian stochastic sampling based on the nested sampling algorithm~\cite{Skilling:2006,Veitch:2009hd}.
Specifically, we use the parameter estimation software, LALInference~\cite{Veitch:2014wba},
which is one of the software programs of LIGO Algorithm Library (LAL) software suite~\cite{LAL}. 
We select low-mass mergers in GWTC-2~\cite{Abbott:2020niy}
which have higher frequency cutoff ($f_\mathrm{high} \gtrsim 120~\mathrm{Hz}$)
and larger inspiral signal-to-noise ratios (SNRs) ($\rho_\mathrm{inspiral} \gtrsim 9$)
(see Table~V in the paper on tests of GR by the LVC~\cite{Abbott:2020jks}).
We take the low-frequency cutoff $f_\mathrm{low}= 20~\mathrm{Hz}$ for all events 
but 30$~\mathrm{Hz}$ for Hanford data for GW170608  
by following the papers which reported the detection~\cite{Abbott:2017gyy}
and the high-frequency cutoff 
$f_\mathrm{high}=150~\mathrm{Hz}$ for GW151226, 
$f_\mathrm{high}=180~\mathrm{Hz}$ for GW170608,
$f_\mathrm{high}=160~\mathrm{Hz}$ for GW190707,
$f_\mathrm{high}=125~\mathrm{Hz}$ for GW190720,
$f_\mathrm{high}=160~\mathrm{Hz}$ for GW190728, and
$f_\mathrm{high}=175~\mathrm{Hz}$ for GW190924,
which are determined to be restricted to the inspiral regime.

\subsection{Source parameters and their priors}
\label{sec:parameters}
The source parameters and their prior probability distributions are basically chosen to follow 
those adopted in the paper on GWTC-2 by the LVC~\cite{Abbott:2020niy} and
our recent work for GW analysis of BNSs~\cite{Narikawa:2018yzt, Narikawa:2019xng}.
We mention the specific choices adopted below.

For BBH hypothesis, the parameters are 
the component masses $m_{1,2}$, where we assume $m_1\geq m_2$;
the orbit-aligned dimensionless spin components of the objects $\chi_{1,2}$; 
the luminosity distance to the source $d_L$; 
the binary inclination angle $\theta_\mathrm{JN}$, 
which is the angle between the total angular momentum and the line of sight; 
the polarization angle $\psi$; 
the coalescence time $t_c$; and the phase at the coalescence time $\phi_c$.
For binary ECO hypothesis, we add the binary tidal deformability $\tilde{\Lambda}$, $\delta\tilde{\Lambda}$, and the SIQMs $\delta\kappa_{1,2}$.

We employ a uniform prior on the detector-frame component 
masses $m_{1,2}^\mathrm{det}$ in the range $[1.0,~60.0]M_\odot$.
We assume a uniform prior on the spin magnitudes $\chi_{1,2}$ in the range $[-0.99,~0.99]$.
We assume a uniform prior on both the binary tidal deformability $\tilde{\Lambda}$ 
and the asymmetric contribution $\delta\tilde{\Lambda}$
in the range $[-3000,~3000]$
and a uniform prior on the SIQMs for individual objects $\delta\kappa_{1,2}$ in the range $[-200,~200]$.
While in the analysis of the paper on tests of GR by the LVC, 
they restrict $\delta\kappa_a=(\delta\kappa_1-\delta\kappa_2)/2=0$, 
implying $\delta\kappa_1=\delta\kappa_2=\delta\kappa_s$~\cite{Abbott:2020jks},
we do not assume so.

\section{Results}
\label{sec:results}

We analyze the BBH events with the low-mass (or long inspiral regime) and higher SNR for the inspiral regime 
among GWTC-2 events~\cite{Abbott:2020niy, GWTC-2}.
Here, we use the public data on the Gravitational Wave Open Science Center
(https://www.gw-openscience.org) released by the LVC.
We analyze each event using inspiral-only template \texttt{TF2g\_Tidal\_SIQM}~waveform model.
First, we present the results for the largest SNR event GW170608 in detail.
Next, we show the results for other events and model selection by the Bayes factor combining six events\footnote{The corrections to the tidal phasing in Eq.~(4) described in the Erratum~\cite{Narikawa_Erratum} are numerically very small. 
The phase difference between the previous expression and the correct one is less than
$\mathcal{O}(10^{-3})$ (rad) up to 1000~Hz for a binary neutron star
with $m_A=1.68~M_\odot$, $m_B=1.13~M_\odot$, $\Lambda_A=102$, and $\Lambda_B=840$.
The difference is less than $\mathcal{O}(10^{-2})$ (rad) up to 200 Hz for a binary exotic compact object with $m_A=13~M_\odot$, $m_B=6.6~M_\odot$, $\Lambda_A=-700$, and $\Lambda_B=-1400$.
Therefore, their impact on the main results is negligible.}.

\begin{figure}[htbp]
  \begin{center}
 \begin{center}
    \includegraphics[keepaspectratio=true,height=80mm]{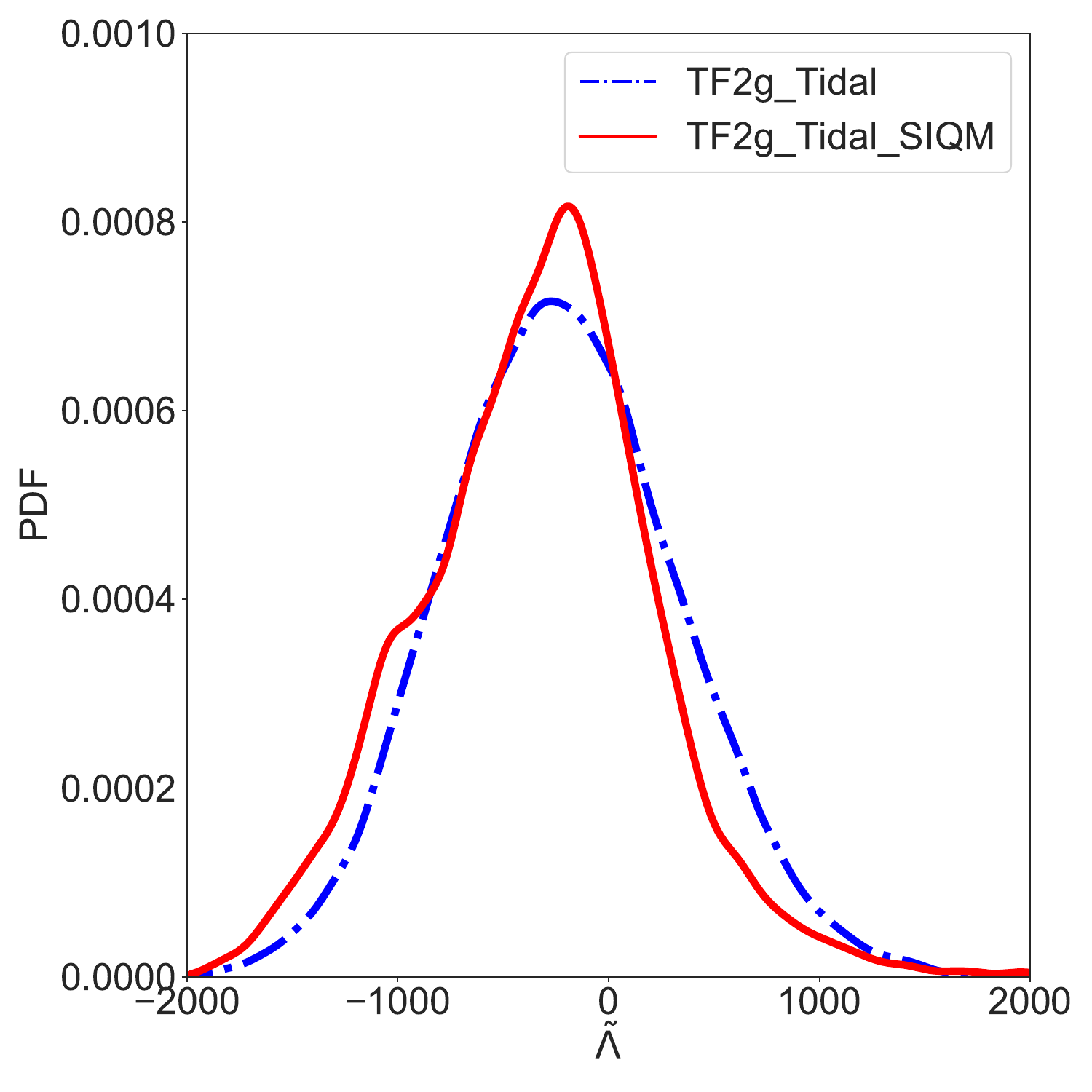}\\
 \end{center}
  \caption{
Marginalized posterior PDFs of $\tilde{\Lambda}$ for low-mass event GW170608
with the \texttt{TF2g\_Tidal}~(blue, dashed) and \texttt{TF2g\_Tidal\_SIQM}~(red, solid) waveform models.
We set $f_\mathrm{high}=180~\mathrm{Hz}$.
Adding the SIQM terms do not affect the constraint on the tidal deformability $\tilde{\Lambda}$.
}%
\label{fig:Lamt_TF2g_GW170608}
\end{center}
\end{figure}

\begin{figure}[htbp]
  \begin{center}
 \begin{center}
    \includegraphics[keepaspectratio=true,height=80mm]{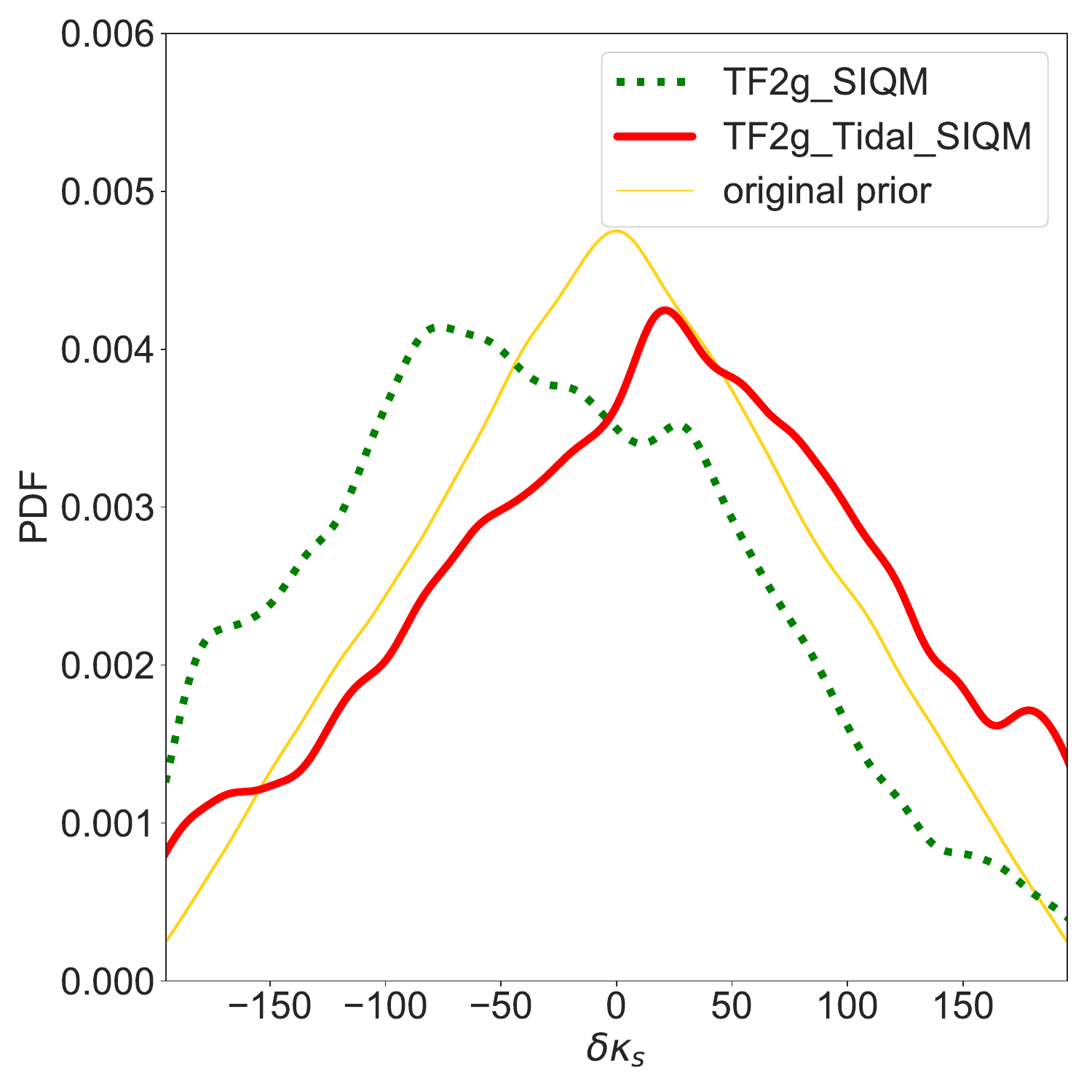}\\
 \end{center}
  \caption{
Marginalized posterior PDFs of $\delta\kappa_s$ for GW170608
using the \texttt{TF2g\_SIQM}~(green, dotted) and \texttt{TF2g\_Tidal\_SIQM}~(red, solid) waveform models.
We set $f_\mathrm{high}=180~\mathrm{Hz}$.
They are weighted by dividing the original prior: uniform on $\delta\kappa_{1,2}$.
The original prior is also shown by solid yellow curve.
The peak of $\delta\kappa_s$ is shifted toward zero by adding the tidal terms.
}%
\label{fig:dkappaS_TF2g_GW170608}
\end{center}
\end{figure}

\begin{figure}[htbp]
  \begin{center}
 \begin{center}
    \includegraphics[keepaspectratio=true,height=80mm]{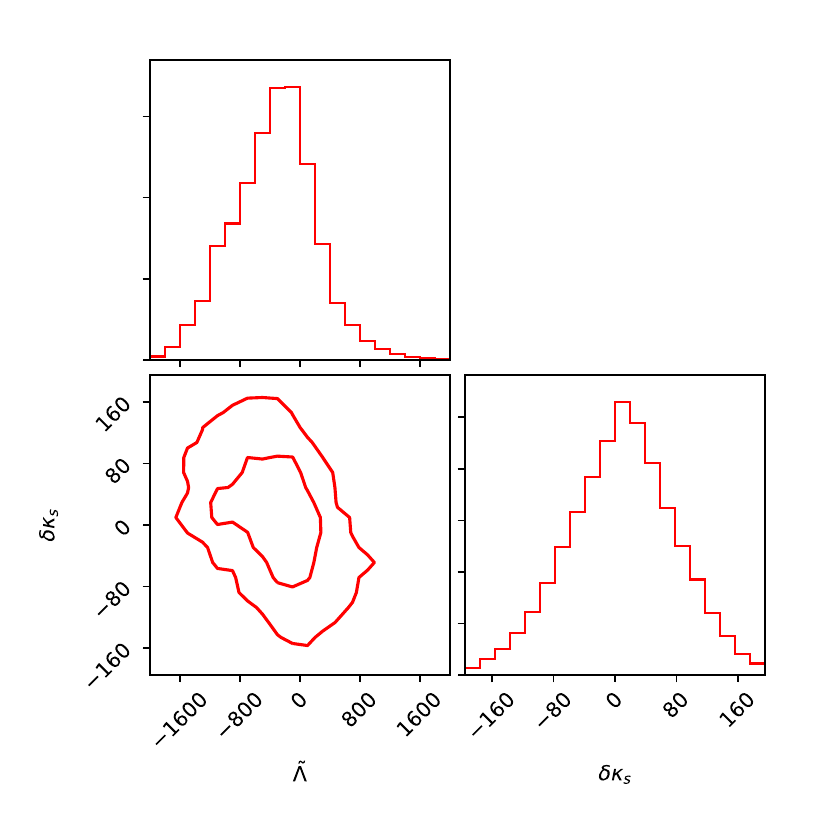}\\
 \end{center}
  \caption{
Corner plot on the $\tilde{\Lambda}$-$\delta\kappa_s$ plane for GW170608
using the \texttt{TF2g\_Tidal\_SIQM}~waveform model
by setting $f_\mathrm{high}=180~\mathrm{Hz}$.
The contours correspond to 50\% and 90\% credible regions.
The constraints show that GW170608 are consistent with BBH in GR ($\tilde{\Lambda} = \delta\kappa_s = 0$).
}%
\label{fig:Lamt-dkappaS_TF2g_GW170608}
\end{center}
\end{figure}

\begin{table}[htbp]
\begin{center}
\caption{
The logarithm of the Bayes factors for a signal compared to Gaussian noise $\log_{10}\mathrm{BF}_{s/n}$ and SNRs 
for GW170608 using the \texttt{TF2g}, \texttt{TF2g\_Tidal}, \texttt{TF2g\_SIQM}, and \texttt{TF2g\_Tidal\_SIQM}~waveform models.
The tidal and SIQM terms do not affect the Bayes factor and SNR.
}
\begin{tabular}{lccccc}
\hline \hline
 & ~~$\log_{10}\mathrm{BF}_{s/n}$ & SNR \\ \hline
\texttt{TF2g}~(BBH in GR) & 71.3 & 14.7 \\
\texttt{TF2g\_Tidal} & 69.8 & 14.7 \\
\texttt{TF2g\_SIQM} & 70.4 & 14.7 \\
\texttt{TF2g\_Tidal\_SIQM} & 69.3 & 14.7 \\
\hline \hline
\end{tabular}
\label{table:BF_TF2g_HFBTidal_SIQM_GW170608}
\end{center}
\end{table}

\subsection{Estimating tidal deformability and SIQMs for GW170608}
\label{sec:GW170608}

We show the results of low-mass presumed BBH event GW170608.
We present posteriors of binary tidal deformability and the SIQM for GW170608.
Figure~\ref{fig:Lamt_TF2g_GW170608} shows posteriors of $\tilde{\Lambda}$ for GW170608.
We set $f_\mathrm{high}=180~\mathrm{Hz}$.
The posterior distribution of $\tilde{\Lambda}$ for the \texttt{TF2g\_Tidal}~waveform model (blue dot-dashed) 
is consistent with the one for the \texttt{TF2g\_Tidal\_SIQM}~waveform model (red solid).
Adding the SIQM terms does not affect the constraint on the tidal deformability $\tilde{\Lambda}$.

Figure~\ref{fig:dkappaS_TF2g_GW170608} shows posteriors of $\delta\kappa_s$ for GW170608 using the \texttt{TF2g\_SIQM}~and \texttt{TF2g\_Tidal\_SIQM}~waveform models.
Here, they are weighted by dividing the original prior, uniform on $\delta\kappa_{1,2}$,
to effectively impose a uniform prior on $\delta\kappa_s$.
We find that $\delta\kappa_s$ is poorly constrained for GW170608 for both templates, 
which is consistent with the results shown for the templates with only the SIQMs in Refs.~\cite{Krishnendu:2019tjp, Abbott:2020jks}.
Equation~(\ref{eq:SIQM}) shows that $Q_{i}=0$ when $\chi_{i}=0$, independent of the value of $\delta\kappa_i$,
and $\delta\kappa_{i}$ is not constrained unless at least one of the magnitude of the component spin $\chi_{i}$ exclude zero.
It is natural that our constraints on $\delta\kappa_s$ are poorer than LVC's 
since we do not assume $\delta\kappa_a=0$.
Comparing the posteriors for the \texttt{TF2g\_SIQM}~(green, dotted) with \texttt{TF2g\_Tidal\_SIQM} ~(red, solid) waveform models,
adding the tidal terms shifts the peak of $\delta\kappa_s$ toward zero.

Figure~\ref{fig:Lamt-dkappaS_TF2g_GW170608} shows the corner plot on the $\tilde{\Lambda}$-$\delta\kappa_s$ plane 
for GW170608 using the \texttt{TF2g\_Tidal\_SIQM}~waveform model,
which shows that the observed data are consistent with BBH in GR 
($\tilde{\Lambda} = \delta\kappa_s =0$).
This figure shows the posterior of $\delta\kappa_s$ for the uniform prior on $\delta\kappa_{1,2}$.
We find a weak negative correlation between $\tilde{\Lambda}$ and $\delta\kappa_s$.

Table~\ref{table:BF_TF2g_HFBTidal_SIQM_GW170608} shows 
the logarithm of the Bayes factor for a signal 
compared to Gaussian noise $\log_{10}\mathrm{BF}_{s/n}$.
Adding the tidal and SIQM terms does not drastically change the Bayes factor $\log_{10}\mathrm{BF}_{s/n}$ and SNRs.


\begin{figure*}[htbp]
  \begin{center}
 \begin{center}
    \includegraphics[keepaspectratio=true,height=180mm]{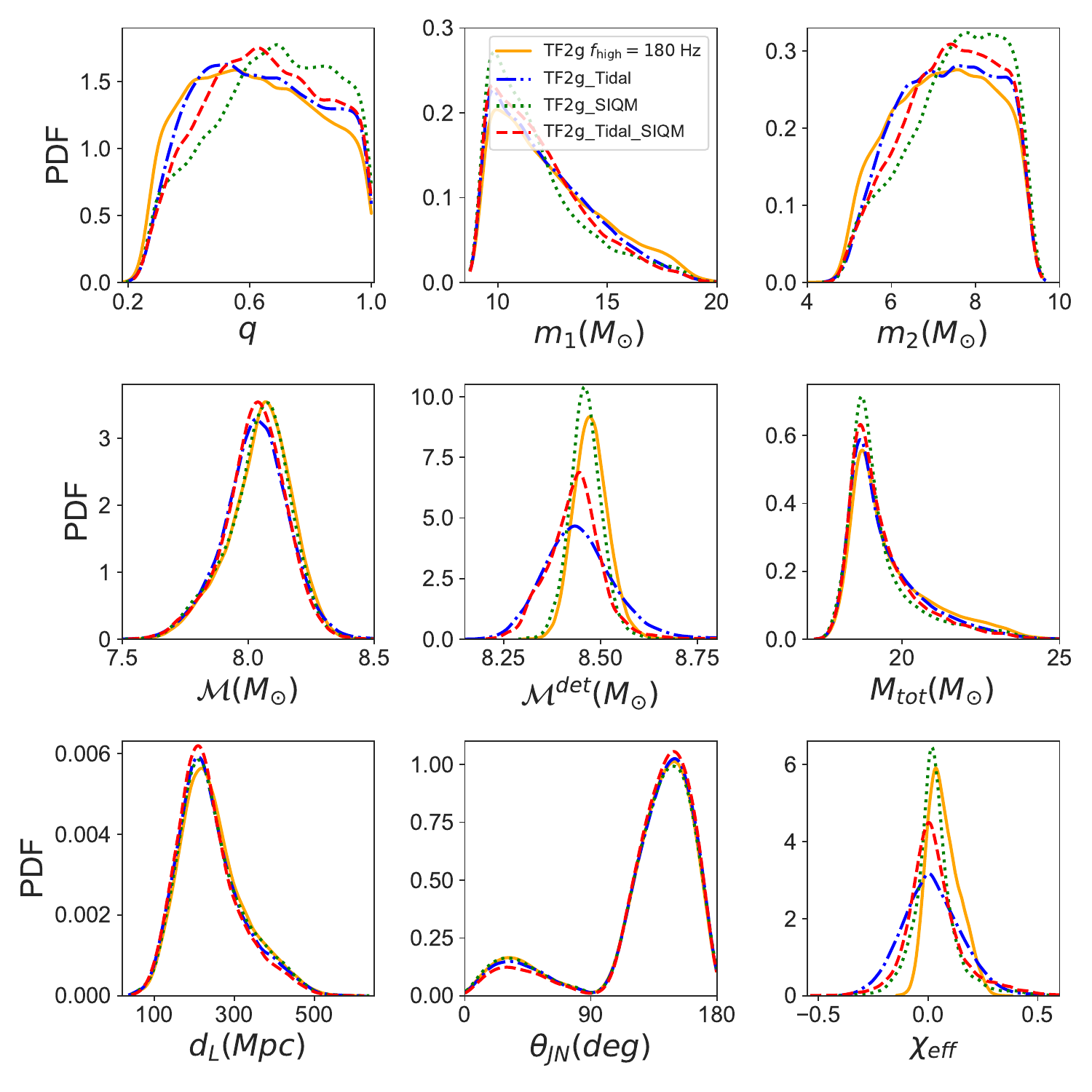}\\
 \end{center}
  \caption{
Marginalized posterior PDFs of various parameters for GW170608 derived by using 
the \texttt{TF2g}~(orange, solid), \texttt{TF2g\_Tidal}~(blue, dot-dashed), \texttt{TF2g\_SIQM}~(green, dotted), and \texttt{TF2g\_Tidal\_SIQM} (red, dashed)~waveform models, 
by setting $f_\mathrm{high}=180~\mathrm{Hz}$.
The top-left, top-middle, top-right, middle-left, center, middle-right, bottom-left, bottom-middle, and bottom-right panels show  the mass ratio $q$, the primary mass $m_1$, the secondary mass $m_2$, the source-frame chirp mass ${\cal M}$, the detector-frame chirp mass $\mathcal{M}^\mathrm{det}$, the total mass $M_\mathrm{tot}$, the luminosity distance to the source $d_L$, the inclination angle $\theta_\mathrm{JN}$, and the effective spin parameter $\chi_\mathrm{eff}$, respectively.
Adding either the tidal or the SIQM terms shifts the peak of $\chi_\mathrm{eff}$ toward zero,
due to the difficulty of measuring $\chi_\mathrm{eff}$.
}%
\label{fig:all_TF2g_Tidal_SIQM_GW170608}
\end{center}
\end{figure*}


\subsection{Biases of binary parameters due to tidal and SIQM terms for GW170608}
\label{sec:bias}

Figure~\ref{fig:all_TF2g_Tidal_SIQM_GW170608} shows 
the marginalized posterior PDFs of parameters other than the tidal deformability and the SIQM 
obtained by using the \texttt{TF2g}~model (orange, solid),
\texttt{TF2g\_Tidal}~(blue, dot-dashed), 
\texttt{TF2g\_SIQM}~(green, dotted),
and \texttt{TF2g\_Tidal\_SIQM}~(red, dashed) waveform models.
We present the distribution of 
the mass parameters: the mass ratio $q=m_2/m_1$, the primary mass $m_1$, the secondary mass $m_2$, 
the chirp mass ${\cal M}$, the detector-frame chirp mass ${\cal M}^\mathrm{det}$ and the total mass $M_\mathrm{tot}$, 
the luminosity distance $d_L$, the binary inclination $\theta_\mathrm{JN}$, 
and the effective inspiral spin $\chi_\mathrm{eff}$,
which gives the leading-order spin effect on the binary phase evolution~\cite{Ajith:2009bn, Santamaria:2010yb}
defined as $\chi_\mathrm{eff}=(m_1 \chi_1 + m_2 \chi_2)/M_\mathrm{tot}$.
The source-frame chirp mass is derived by assuming the Hubble constant $H_0 = 69~\mathrm{km}~\mathrm{s}^{-1}~\mathrm{Mpc}^{-1}$ (a default value in LALInference adopted from Planck 2013 results~\cite{Ade:2013sjv}).
$q$, $m_1$, $m_2$, ${\cal M}$, $M_\mathrm{tot}$, $d_\mathrm{L}$, and $\theta_\mathrm{JN}$
are not affected by adding the tidal and SIQM terms to the \texttt{TF2g}~baseline model.
Adding either the tidal or the SIQM terms shifts the peak of $\chi_\mathrm{eff}$ toward zero, 
as also shown in Fig.~5 of Ref.~\cite{Krishnendu:2019tjp} for SIQM,
which is because measurement of $\chi_\mathrm{eff}$ becomes difficult.
This also biases the estimate of ${\cal M}^\mathrm{det}$.
Comparing the posteriors of $\chi_\mathrm{eff}$ for the \texttt{TF2g\_Tidal}~(blue dot-dashed) 
with \texttt{TF2g\_Tidal\_SIQM}~(red solid) waveform models,
adding the SIQM terms reduces the statistical error of $\chi_\mathrm{eff}$.

\subsection{Constraints on tidal deformability and SIQMs for GWTC-2 events}
\label{sec:GWTC2}

\begin{figure*}[htbp]
  \begin{center}
\begin{tabular}{cc}
 \begin{minipage}[b]{0.45\linewidth}
 \begin{center}
    \includegraphics[keepaspectratio=true,height=80mm]{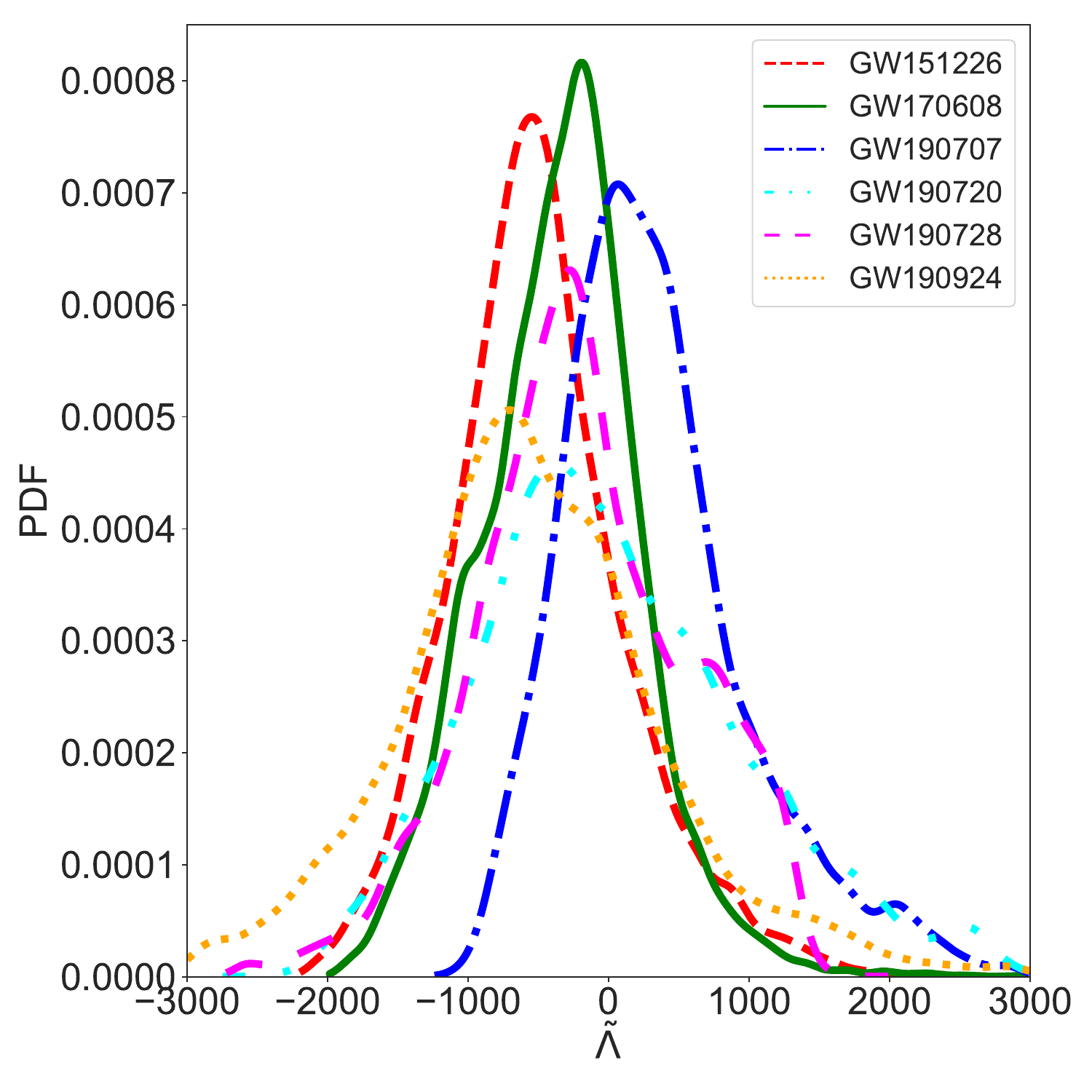}\\
 \end{center}
 \end{minipage}
 \begin{minipage}[b]{0.45\linewidth}
  \begin{center}
    \includegraphics[keepaspectratio=true,height=80mm]{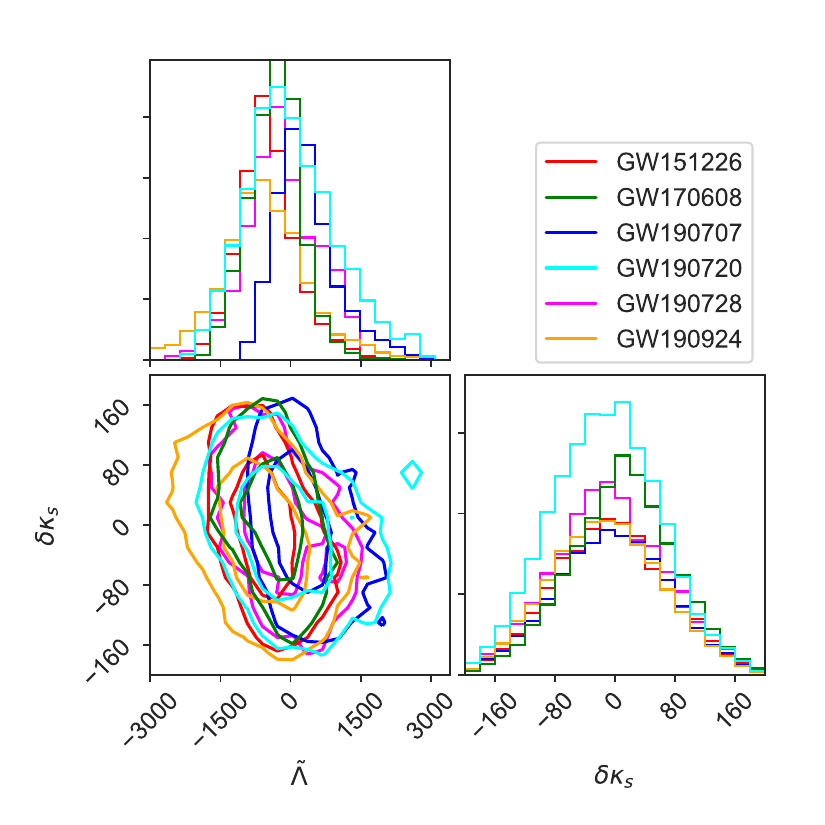}\\
 \end{center}
 \end{minipage}
\end{tabular}
  \caption{
Left: the same as Fig.~\ref{fig:Lamt_TF2g_GW170608} but for GW151226 (red dashed), GW170608 (green solid), 
GW190707 (blue dot-dashed), GW190720 (cyan dash-dot-dotted), GW190728 (magenta loosely dashed),
and GW190924 (orange dotted),
using the \texttt{TF2g\_Tidal\_SIQM}~waveform model 
by setting $f_\mathrm{high}=150$, 180, 160, 125, 160, and 175$~\mathrm{Hz}$, respectively.
These constraints show that all events are consistent with BBH in GR ($\tilde{\Lambda}=0$).
Right: the same as Fig.~\ref{fig:Lamt-dkappaS_TF2g_GW170608} but for GW151226 (red), GW170608 (green), 
GW190707 (blue), GW190720 (cyan), GW190728 (magenta), and GW190924 (orange)
using the \texttt{TF2g\_Tidal\_SIQM}~waveform model for $f_\mathrm{high}=150$, 180, 160, 125, 160, and 175$~\mathrm{Hz}$, respectively.
These constraints show that all events are consistent with BBH in GR ($\tilde{\Lambda} = \delta\kappa_s = 0$).
}%
\label{fig:Lamt-dkappaS_TF2g_Sum}
\end{center}
\end{figure*}

We also analyzed other five low-mass events: 
GW151226 with $f_\mathrm{high}=150~\mathrm{Hz}$, 
GW190707 with $f_\mathrm{high}=160~\mathrm{Hz}$,
GW190720 with $f_\mathrm{high}=125~\mathrm{Hz}$,
GW190728 with $f_\mathrm{high}=160~\mathrm{Hz}$, and
GW190924 with $f_\mathrm{high}=175~\mathrm{Hz}$,
using the \texttt{TF2g\_Tidal\_SIQM}~waveform model.
We present posteriors of binary tidal deformability and the SIQMs for GW151226, GW170608, GW190707, 
GW190720, GW190728, and GW190924.
The left panel of Fig.~\ref{fig:Lamt-dkappaS_TF2g_Sum} shows posteriors of $\tilde{\Lambda}$ for six events
using the \texttt{TF2g\_Tidal\_SIQM}~waveform model.
This figure shows the posterior of $\delta\kappa_s$ for the uniform prior on $\delta\kappa_{1,2}$.
The 90\% symmetric credible ranges of $\tilde{\Lambda}$ are summarized in Table~\ref{table:BF_TF2g_HFBTidal_SIQM_Sum},
which are 
$[-1441,~649]$ for GW151226,
$[-1265,~565]$ for GW170608,
$[-1265,~565]$ for GW170608,
$[-590,~1661]$ for GW190707, 
$[-1445,~1768]$ for GW190720,
$[-1432,~1078]$ for GW190728, and 
$[-2041,~1118]$ for GW190924.

The right panel of Fig.~\ref{fig:Lamt-dkappaS_TF2g_Sum} shows the posterior distribution on the $\tilde{\Lambda}$-$\delta\kappa_s$ plane.
For all events, there exists negative correlation between $\tilde{\Lambda}$ and $\delta\kappa_s$.
We find that $\delta\kappa_s$ is poorly constrained for all events we analyzed, which is consistent with the results in Refs.~\cite{Krishnendu:2019tjp, Abbott:2020jks}.
We do not show the credible intervals for $\delta\kappa_s$, 
since all the posterior PDFs present tails reaching the edge of the prior
when they are divided by the original prior (uniform on $\delta\kappa_{1,2}$)
as shown in Fig.~\ref{fig:dkappaS_TF2g_GW170608} for GW170608
and in Refs.~\cite{Krishnendu:2019tjp, Abbott:2020jks} for other cases.
It is natural that our constraints on $\delta\kappa_s$ are poorer than LVC's, 
since we do not assume $\delta\kappa_a=0$, while LVC did.

\subsection{Model selection between ECO and BBH}
\label{sec:ModelSelection}

The Bayes factor between the binary ECO and BBH waveform model, $\mathrm{BF}^\mathrm{ECO}_\mathrm{BBH}$, 
quantifies the statistical significance 
of one hypothesis over the other.
The binary ECO hypothesis (the \texttt{TF2g\_Tidal\_SIQM}~waveform model) 
is disfavored compared to BBH (the \texttt{TF2g}~waveform model) for all events as shown Table~\ref{table:BF_TF2g_HFBTidal_SIQM_Sum}\footnote{$\log_{10}\mathrm{BF} > 1.5$ is often interpreted as a strong preference for one model over another, 
and $\log_{10}\mathrm{BF} > 2$ is interpreted as decisive evidence~\cite{Jeffreys:1961}.}.

The combined Bayes factor of six events estimated by using Eq.~(\ref{eq:combinedBF}),
$\log_{10}\mathrm{BF}^\mathrm{ECO}_\mathrm{BBH,total}=-10.4$,
shows that the BBH hypothesis is preferred to the binary ECO hypothesis.

\begin{table}[htbp]
\begin{center}
\caption{
The 90\% symmetric credible ranges of $\tilde{\Lambda}$,
the logarithm of the Bayes factors between binary ECO and BBH, i.e., 
$\log_{10}\mathrm{BF}^\mathrm{ECO}_\mathrm{BBH}$,
and SNRs for GW151226, GW170608, GW190707, GW190720, GW190728, GW190924, 
and the result combining six events
using the \texttt{TF2g\_Tidal\_SIQM}~waveform model with respective $f_\mathrm{high}$.
For individual events, $\log_{10}\mathrm{BF}^\mathrm{ECO}_\mathrm{BBH}$ are negative, 
thus favoring the BBH in GR compared to binary ECO.
For the combined case, $\log_{10}\mathrm{BF}^\mathrm{ECO}_\mathrm{BBH,total}$ is also negative, 
thus disfavoring the binary ECO.
}
\begin{tabular}{lccccc}
\hline \hline
Event & ~~$f_\mathrm{high}~(\mathrm{Hz})$ & ~~$\tilde{\Lambda}$ & ~~$\log_{10}\mathrm{BF}^\mathrm{ECO}_\mathrm{BBH}$ & ~~SNR \\ \hline
GW151226 & 150 & $[-1441,~649]$ & -0.45 & 10.7 \\
GW170608 & 180 & $[-1265,~565]$ &  -2.1 & 14.7 \\
GW190707 & 160 & $[ -590,~1661]$ & -2.1 & 11.2 \\
GW190720 & 125 & $[-1445,~1768]$ & -1.8 & 9.3 \\
GW190728 & 160 & $[-1432,~1078]$ & -2.0 & 12.1 \\
GW190924 & 175 & $[-2041,~1118]$ & -2.0 & 11.4 \\
Combined   &  -     & -     & -10.4 & - \\
\hline \hline
\end{tabular}
\label{table:BF_TF2g_HFBTidal_SIQM_Sum}
\end{center}
\end{table}

\section{Conclusion}
\label{sec:summary}
We implemented the tidal and SIQM terms
in the aligned-spin PN inspiral waveform model \texttt{TF2g}, which we call \texttt{TF2g\_Tidal\_SIQM}.
We analyzed six low-mass events GW151226, GW170608, GW190707, 
GW190720, GW190728, and GW190924
using the \texttt{TF2g\_Tidal\_SIQM}~waveform model as templates.
The obtained results are the first constraints on the tidal deformability of 
the events classified as BBH in GWTC-2 events, motivated by ECO hypotheses.
We found that all events that we have analyzed are consistent with BBH mergers in GR.
The logarithmic Bayes factor $\log_{10}\mathrm{BF}^\mathrm{ECO}_\mathrm{BBH}$ for individual events is 
less than -1.5 except for GW151226,
thus favoring the BBH in GR compared to binary ECO by Bayesian model selection.
The combined logarithmic Bayes factor between the binary ECO and BBH 
is $\log_{10}\mathrm{BF}^\mathrm{ECO}_\mathrm{BBH,total}=-10.1$, which means that the ECO or non-GR model (with Tidal and SIQM parameters) is also disfavored compared to BBH in GR.

In this paper, we used the inspiral-only waveform model as templates 
because there is no robust prediction waveform 
of the merger-ringdown regimes of the binary ECO merger.
It might be interesting to use a toy model for postcontact regimes of binary ECOs, 
which has been recently derived~\cite{Toubiana:2020lzd}.
Also, GW echoes could be used to examine postmerger regime.
Such extensions including the waveform after the inspiral regime would allow us to analyze heavy-mass events and 
to put a different type of constraints on ECO hypothesis.

\section*{Acknowledgment}
T. Narikawa thanks 
Chris van den Broeck and Anuradha Samajdar for useful discussions, and he is also thankful for the hospitality of the Chris’s group,
in particular Khun Sang Phukon, Archisman Gosh, and Tim Dietrich, during his stay at Nikhef. 
We thank Aditya Vijaykumar, Nathan K. Johnson-McDaniel, Rahul Kashyap, Arunava Mukherjee, and Parameswaran Ajith for sharing and discussing their results on a similar study~\cite{Aditya:2021}.
We would like to thank Soichiro Morisaki, Kyohei Kawaguchi, and Hideyuki Tagoshi for fruitful discussions.
We thank Quentin Henry and Luc Blanchet for pointing out the complete GW tidal phase they have derived
and confirming the correction that the expression that we have rewritten with their value for the mass
quadrupole part as a function of $\Lambda_{1,2}$, especially the 2 and 2.5 PN-order terms.
T. Narikawa was supported in part by a Grant-in-Aid for JSPS Research Fellows.
This work is supported by Japanese Society for the Promotion of Science (JSPS) KAKENHI Grants
No.~JP21K03548, No.~JP17H06361, No.~JP17H06358, No.~JP17H06357, and No.~JP20K03928.
We would also like to thank Computing Infrastructure ORION in Osaka City University. 
We are also grateful to the LIGO-Virgo
Collaboration for the public release of gravitational-wave data of
GW151226, GW170608, GW190707\_093326, and GW190924\_021846. 
This research has made use of data, software, and web tools
obtained from the Gravitational Wave Open Science Center
(https://www.gw-openscience.org), a service of LIGO Laboratory, the LIGO
Scientific Collaboration and the Virgo Collaboration. LIGO is funded by
the U. S. National Science Foundation. Virgo is funded by the French
Centre National de la Recherche Scientifique (CNRS), the Italian
Istituto Nazionale di Fisica Nucleare (INFN), and the Dutch Nikhef, with
contributions by Polish and Hungarian institutes.

\appendix

\section{Results by using \texttt{TF2}~waveform model}
\label{sec:results_TF2}

While we show the results by using the \texttt{TF2g\_Tidal\_SIQM}~waveform model in Sec.~\ref{sec:results},
we show the results for seven events added GW190814 
by using the \texttt{TF2\_Tidal\_SIQM}~waveform model in this Appendix.
For highly unequal mass ratio events GW190814, the \texttt{TF2g}~waveform model does not work well
due to setting the uncalculated terms at high PN order to zero.
We take the low-frequency cutoff $f_\mathrm{low}= 20~\mathrm{Hz}$ for Hanford data 
and 30$~\mathrm{Hz}$ for Livingston data for GW190814 
by following the papers which reported the detection~\cite{Abbott:2020khf}
and the high-frequency cutoff $f_\mathrm{high}=140~\mathrm{Hz}$.

We present posteriors of binary tidal deformability and the SIQMs for GW151226, GW170608, GW190707, 
GW190720, GW190728, GW190814 and GW190924 by using the \texttt{TF2\_Tidal\_SIQM}~waveform model.

\begin{figure*}[htbp]
  \begin{center}
\begin{tabular}{cc}
 \begin{minipage}[b]{0.45\linewidth}
 \begin{center}
    \includegraphics[keepaspectratio=true,height=80mm]{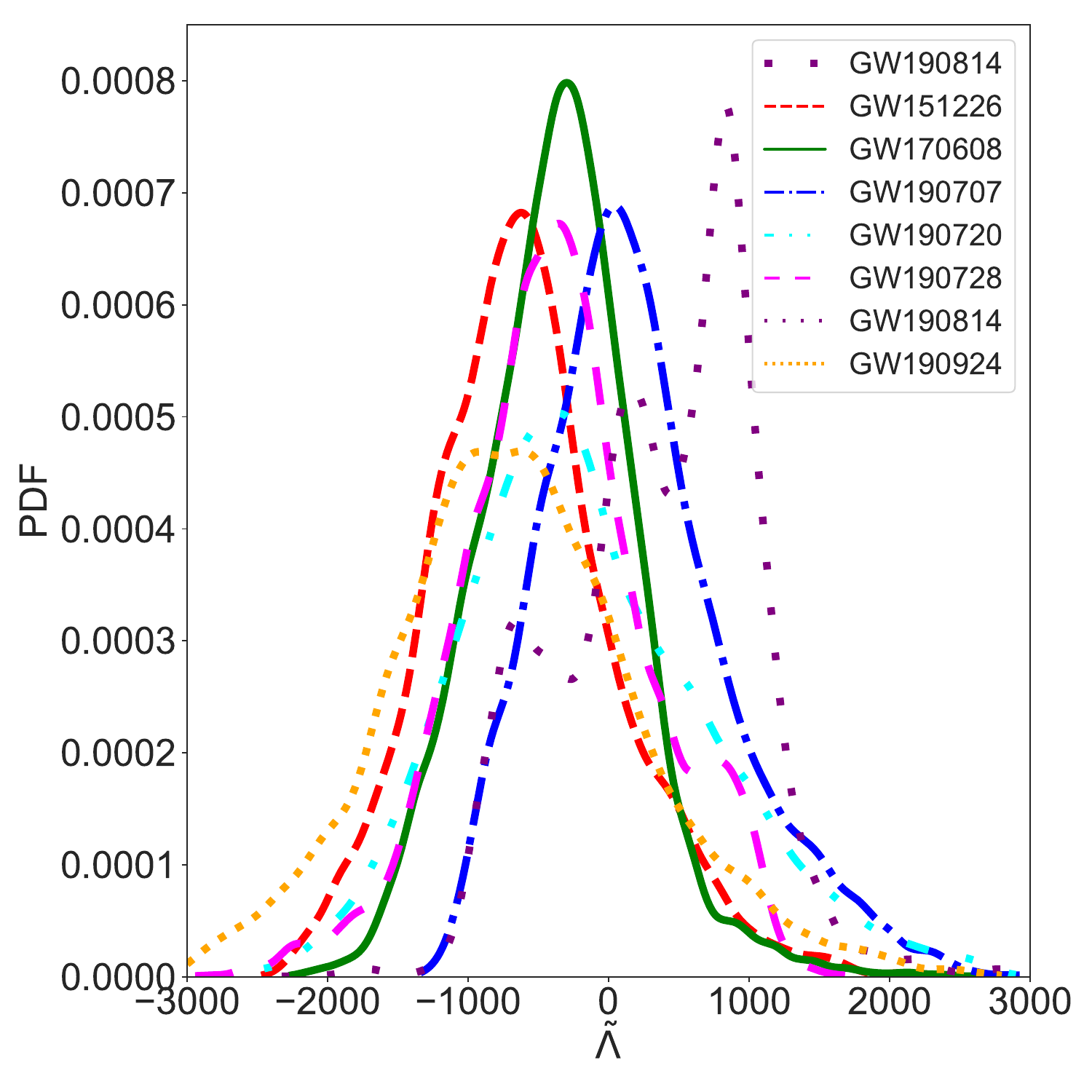}\\
 \end{center}
 \end{minipage}
 \begin{minipage}[b]{0.45\linewidth}
  \begin{center}
    \includegraphics[keepaspectratio=true,height=80mm]{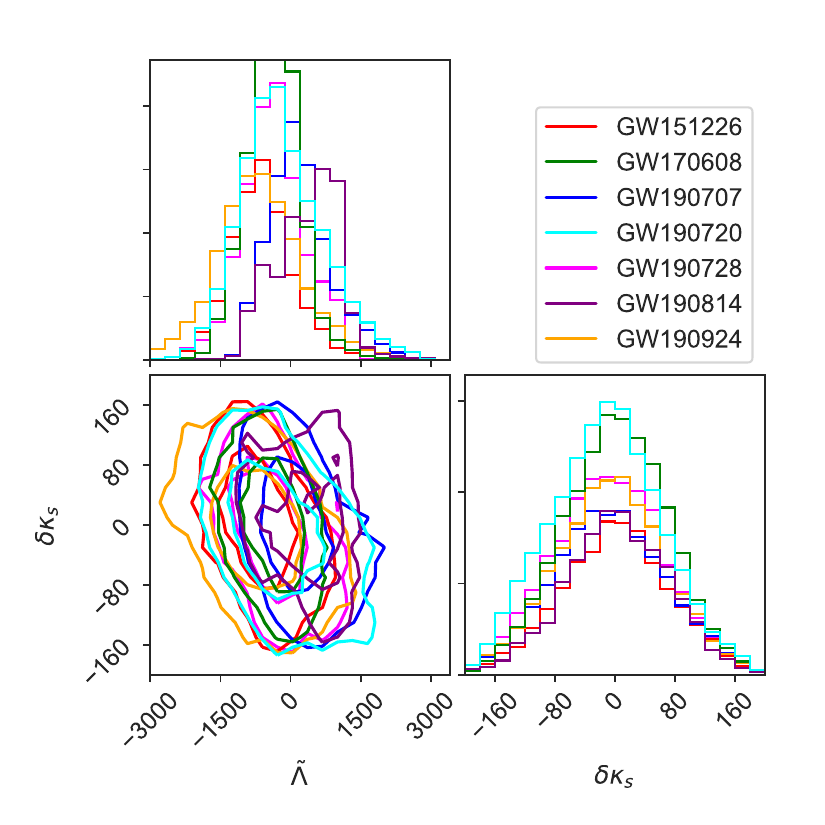}\\
 \end{center}
 \end{minipage}
\end{tabular}
  \caption{
Left: the same as 
the left panel of Fig.~\ref{fig:Lamt-dkappaS_TF2g_Sum} but for the \texttt{TF2\_Tidal\_SIQM}~waveform model
and GW190814 (purple loosely dotted) is added. 
These constraints show that all events are consistent with BBH in GR ($\tilde{\Lambda}=0$).
Right: the same as 
the right panel of Fig.~\ref{fig:Lamt-dkappaS_TF2g_Sum}
but for the \texttt{TF2\_Tidal\_SIQM}~waveform model
and GW190814 (purple) is added.
These constraints show that all events are consistent with BBH in GR ($\tilde{\Lambda} = \delta\kappa_s = 0$).
}%
\label{fig:Lamt-dkappaS_TF2_Sum}
\end{center}
\end{figure*}

The left panel of Fig.~\ref{fig:Lamt-dkappaS_TF2_Sum} shows posteriors of $\tilde{\Lambda}$ for seven events
using the \texttt{TF2\_Tidal\_SIQM}~waveform model.
%
The right panel of Fig.~\ref{fig:Lamt-dkappaS_TF2_Sum} shows the posterior distribution on $\tilde{\Lambda}$-$\delta\kappa_s$ plane.
The 90\% symmetric credible ranges of $\tilde{\Lambda}$ are summarized in Table~\ref{table:BF_TF2_HFBTidal_SIQM_Sum},
which are 
$[-1656,~590]$ for GW151226,
$[-1299,~520]$ for GW170608,
$[-791,~1487]$ for GW190707, 
$[-1523,~1436]$ for GW190720,
$[-1445,~846]$ for GW190728, 
$[-786,~1272]$ for GW190814, and 
$[-2093,~939]$ for GW190924.
Except for GW190814, 
systematic uncertainty between \texttt{TF2g\_Tidal\_SIQM}~and \texttt{TF2\_Tidal\_SIQM}~results
remains subdominant to statistical uncertainty.

We show the Bayes factor between the binary ECO and BBH waveform model, $\mathrm{BF}^\mathrm{ECO}_\mathrm{BBH}$ in Table~\ref{table:BF_TF2_HFBTidal_SIQM_Sum}.
The binary ECO hypothesis (the \texttt{TF2\_Tidal\_SIQM}~waveform model) 
is disfavored compared to BBH (the \texttt{TF2g}~waveform model) for all events.
The combined Bayes factor of seven events are 
$\log_{10}\mathrm{BF}^\mathrm{ECO}_\mathrm{BBH,total}=-14.0$
shows that BBH hypothesis is preferred to the binary ECO hypothesis 
as shown for \texttt{TF2g\_Tidal\_SIQM}.

\begin{table}[htbp]
\begin{center}
\caption{
The same as Table \ref{table:BF_TF2g_HFBTidal_SIQM_Sum} but for the \texttt{TF2\_Tidal\_SIQM}~waveform model
and GW190814 is added. 
For both the individual events and the combined case, the log Bayes factor 
$\log_{10}\mathrm{BF}^\mathrm{ECO}_\mathrm{BBH,total}$ is negative, 
thus disfavoring binary ECO.
}
\begin{tabular}{lccccc}
\hline \hline
Event & ~~$f_\mathrm{high}~(\mathrm{Hz})$ & ~~$\tilde{\Lambda}$ & ~~$\log_{10}\mathrm{BF}^\mathrm{ECO}_\mathrm{BBH}$ & ~~SNR \\ \hline
GW151226 & 150 & $[-1656,~590]$ & -0.66 & 10.7 \\
GW170608 & 180 & $[-1299,~520]$ &  -1.8 & 14.7 \\
GW190707 & 160 & $[-791,~1487]$ & -2.0 & 11.2 \\
GW190720 & 125 & $[-1523,~1436]$ & -1.3 & 9.3 \\
GW190728 & 160 & $[-1445,~846]$ & -1.9 & 12.1 \\
GW190814 & 140 & $[-786,~1272]$ & -4.0 & 22.0 \\
GW190924 & 175 & $[-2093,~939]$ & -2.2 & 11.4 \\
Combined   &  -     & -     & -14.0 & - \\
\hline \hline
\end{tabular}
\label{table:BF_TF2_HFBTidal_SIQM_Sum}
\end{center}
\end{table}

  

\end{document}